\documentstyle[epsfig,epsf]{article}
\begin{document}

\begin{center}
{\Large {\bf Nuclear Track Detectors for Environmental Studies and Radiation Monitoring }}
\end{center}

\vskip .7 cm

\begin{center}

S. Manzoor$^{a,b,}$\footnote{Corresponding author, manzoor@bo.infn.it},
S. Balestra$^a$,  
M. Cozzi$^a$, 
M. Errico$^a$, 
G. Giacomelli$^a$,  
M. Giorgini$^a$,  
A. Kumar$^{a,c}$,\par     
A. Margiotta$^a$,  
E. Medinaceli$^a$,  
L. Patrizii$^a$,  
V. Popa$^{a,d}$,  
I.E. Qureshi$^b$, 
and V. Togo$^a$
 \par~\par

{\it  $^a$Phys. Dept. of the University of Bologna and INFN, Sezione di 
Bologna, Viale C. Berti Pichat 6/2, I-40127 Bologna, Italy \\ 
\it $^b$PRD, PINSTECH, P.O. Nilore, Islamabad, Pakistan \\
\it $^c$Dept. Of Physics, Sant Longowal Institute of Eng. and Tech., Longowal 
148 106 India \\
\it $^d$Institute of Space Sciences, Bucharest R-77125, Romania} 

\par~\par
{\footnotesize Presented at the 10$^{th}$ Topical Seminar on Innovative Particle and Radiation Detectors, 1-5 October 2006, Siena, Italy.}
\end{center}

\vskip .7 cm
{\bf Abstract.} \normalsize Several improvements were made for Nuclear Track Detectors 
(NTDs) used for environmental studies and for particle searches. A new method 
was used to determine the bulk etch rate of CR39 and Makrofol NTDs. 
It is based on the simultaneous measurement of the diameter and of the height 
of etch-pit cones caused by relativistic heavy ions (158 A GeV 
Pb$^{82+}$ and In$^{49+}$ ions) and their fragments. The use of alcohol in the 
etching solution improves the surface quality of NTDs and it raises 
their thresholds. The detectors were used for the determination of nuclear 
fragmentation cross sections of Iron and Silicon ions of 1.0 and 0.41 
GeV/nucleon. These measurements are important for the determination 
of doses in hadron therapy and for doses received by astronauts. 
The detectors were also used in the search of massive particles 
in the cosmic radiation, for the determination of the mass spectrum 
of cosmic rays and for the evaluation of Po$^{210}$ $\alpha$-decay and of 
natural radon concentrations.

\vspace{5mm}

\large
\section{Introduction}\label{sec:intro} Nuclear Track Detectors (NTDs) 
were used to search for magnetic monopoles, nuclearites and nuclear 
fragments with fractional charges [1-3]. They were also been used for the 
determination of nuclear fragmentation cross-sections [4-6], the measurement 
of the primary cosmic ray composition [7] and radon measurements [8]. 
Most studies were performed with CR39 NTDs, which have thresholds of 
Z/$\beta \geq 5$. \par
In this paper are described the most favourable etching conditions to obtain 
the best surface quality and reduce the number of background tracks in the 
CR39 and Makrofol NTD's used in the Search for Intermediate Mass Magnetic 
monopoles in SLIM and in other experiments. Formerly we used aqueous 
solutions of NaOH and KOH [9-11]. The addition of ethyl alcohol in the 
etchant improves the etched surface quality, reduces the number of surface 
defects and background tracks at the expense of a higher detection threshold. 
We studied in detail the etching conditions adding different percentages 
of ethyl alcohol. Special care was also paid to proper stirring and 
temperature control of the solutions.\par
For the study of the response of NTDs we used beams of 158 A GeV 
Lead and Indium ions at CERN, 1 A GeV Fe$^{26+}$ and Si$^{14+}$ ions at BNL, 
USA and 0.41 A GeV Fe$^{26+}$ at HIMAC, Japan.\par
        The NTDs were then used for the determination of fragmentation 
cross sections and for the detection of environmental contaminations [12].

\section{Experimental} After exposures the detectors were etched in 
temperature controlled etching baths (to $\pm$ 0.1 $^\circ C$). In order 
to have a 
homogeneous solution during the etching and to avoid the deposit of 
etched products on the detector surfaces, the stirring was kept constant 
during the whole etching cycle. \par
The etchants used were water solutions of 6N NaOH and 6N KOH with different 
fractions of ethyl alcohol. These are called ``soft" etching conditions. 
We also used ``strong" etching conditions in order to fastly reduce 
the thickness of the detectors. For this case we used mainly 8N NaOH, 7N KOH, 
8N KOH water solutions. In these conditions many background tracks of 
10 to 17 $\mu$m range were found; they were probably due to carbon, oxygen and proton 
recoils produced in the interactions of neutrons. Moreover the surface quality was not very good. \par
Recently we added alcohol in the following conditions:\par
{\it Soft etching}: For Makrofol we used a 6N KOH solution with 20$\%$ ethyl alcohol by volume at 50 $^\circ C$. For CR39 we used 6N NaOH and KOH at 70 and 
60 $^\circ C$ with 1, 2 and 3$\%$ ethyl alcohol. \par
After etching, the detectors were cleaned and dried in air and the etched 
tracks were measured with the ELBEK automatic measuring system [13]. \par
The presence of alcohol polishes the detector surfaces, improves the transparency of the post-etched detectors and increases the bulk etching velocity.\par
{\it Strong etching}: The Makrofol sheets were etched in 6N KOH solutions with 
20$\%$ ethyl alcohol at 65 $^\circ C$ for 6 h. The transparency of the detectors was good enough for scanning with a stereo microscope. With this etching condition the thickness was reduced from 500 $\mu$m to about 220 $\mu$m. \par
The CR39 sheets from ``wagons" exposed at the high altitude Chacaltaya lab. 
(5230 m a.s.l.) were strongly etched in a solution 8N KOH at 75 $^\circ C$ 
with 1.5$\%$ 
ethyl alcohol for 30 h. \par
For soft etching the CR39 threshold is at Z/$\beta \sim$ 6-7; for strong 
etching it is at Z/$\beta \sim$17-19.

\section{Calibration}The bulk-etch rate $v_B$ was previously determined by 
measuring the mean thickness difference before and after etching; 
we used an electronic depth-measuring instrument with a 1 $\mu$m accuracy. \par
For relativistic charged particles the track etch rate $v_T$ can be considered 
constant. We measured $v_B$ by another method; for normally incident 
particles, the measurable quantities are the cone base diameter D, and the 
height $L_e$, see Fig. 1.\par
The following two relations hold: 

\begin{equation}
L_e = (v_T-v_B)t~~ ,~~ D= 2 v_B t \sqrt { \frac{(v_T-v_B)}{(v_T+v_B)} } 
\end{equation}

From the above two relations the solution for $v_B$ is

\begin{equation}
v_B= \frac{D^2}{4t L_e} \left[ 1+ \sqrt{ 1+ \frac{4L_e^2}{D^2}} ~ \right]
\end{equation}

Relations (1-2) were tested with relativistic Pb and In ions and their 
fragments in both CR39 and Makrofol NTDs. We selected only those tracks for 
which precise measurements of the cone height and diameter was possible. 
Then, using eq. (2) we computed the bulk-etch rate.\par

\begin{figure}[!ht]
\begin{center}
\mbox{\epsfig{figure=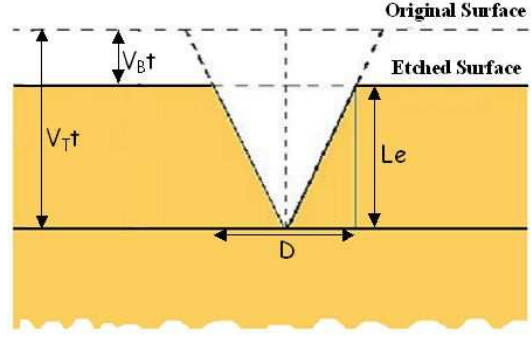,height=4.5 cm}}
\caption{Sketch of an ``etched track" in one side of the detector for a 
normally incident ion. }
\label{fig:1}
\end{center}
\end{figure}

\begin{figure}
\begin{center}
\mbox{\epsfig{figure=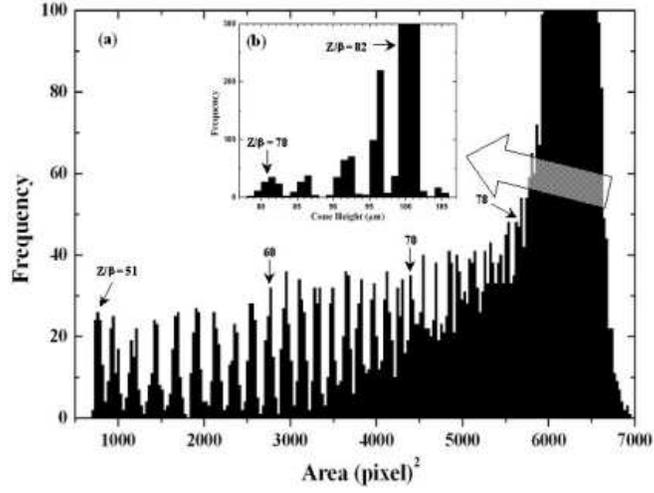,height=6.5cm}}
\caption{(a) Base area distribution of  etched cones in Makrofol from 158 A 
GeV $Pb^{82+}$ ions and their fragments (averages of 2 front face measurements); (b) cone height distribution for $78 \leq Z/ \beta \leq 83$. }
\label{fig:2}
\end{center}
\end{figure}

After ``soft" etching of the CR39 and Makrofol sheets the cone base areas of 
the projectile ions and their fragments were measured with the ELBEK system. 
Fig. 2a shows the base area distribution of Pb ions and their fragments in 
Makrofol. The peaks are well separated from Z/$\beta \sim$51 to 77. 
The charge resolution close to the Pb peak was improved by measuring the heights of the etch-pit cones (see the inset of Fig. 2b). For each detected nuclear fragment we computed the REL and the reduced etch rate 
p=$v_T$/$v_B$ using the formula [14-15]

\begin{equation}
p= \frac{1+(D/2v_B t)^2}{1-(D/2v_B t)^2}
\end{equation}

where 'D' is the diameter of the track, $v_T$ and $v_B$ are the track and 
bulk etch velocities and t is the etching time.

\begin{figure}
\begin{center}
\mbox{\epsfig{figure=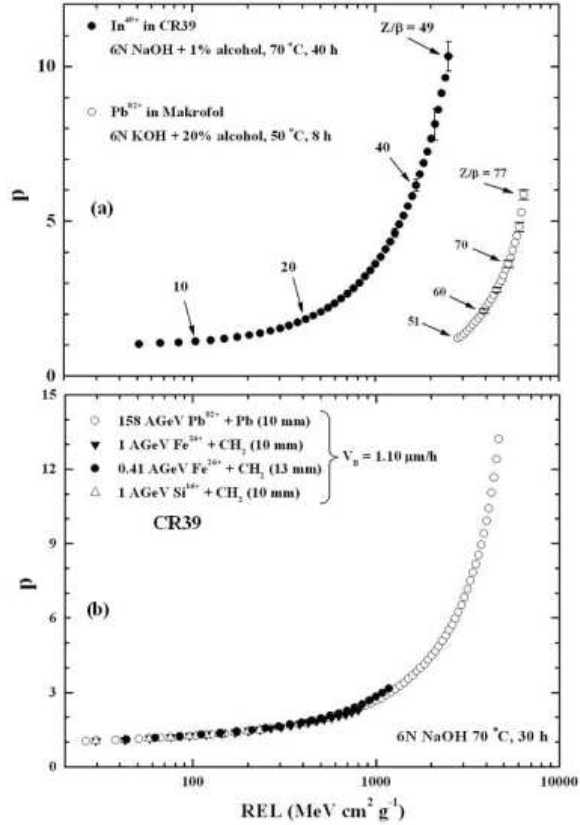,height=11cm}}
\caption{(a) p vs. REL for CR39 and Makrofol NTDs; the points are the data; 
typical statistical standard deviations are shown; for the other data the 
errors are inside the data points. (b) p vs. REL for CR39 exposed to energetic 
Lead, Iron and Silicon ions. The calibrations were made using the new 
method for $v_B$.  }
\label{fig:3}
\end{center}
\end{figure}

The reduced etch rate p versus REL is plotted in Figures 3a,b for CR39 
and Makrofol. The detection thresholds are at REL $\sim$ 50 and 2500 MeV 
cm$^2$ $g^{-1}$ for CR39 and Makrofol detectors, respectively (Fig. 3a). 
A unique calibration curve was obtained for different high-energy heavy 
ions in CR39 (Fig. 3b).

\section{Total Charge Changing Cross Sections}For the determination of the 
total charge changing cross sections, $\sigma_{tot}$, we used beams of 158 A GeV Pb$^{82+}$ 
ions, 1 A GeV Fe$^{26+}$ and Si$^{14+}$ and 0.41 A GeV Fe$^{26+}$ on different 
targets. We measured the number of incident and survived ions before and 
after each target material. \par
The fragmentation charge-changing cross section was evaluated using the formula

\begin{equation}
\sigma_{tot(exp)} = X_T \cdot ln(N_i~ / ~N_s) 
\end{equation}  

where $X_T = A_T/ \rho_T \cdot t_T \cdot N_A$; $N_i$ 
is the number of primary ions, $N_S$ the number of survived ions after the 
target. $A_T$, $\rho_T$ and $t_T$ are the mass number, density and thickness of the target material. \par
The cross section on hydrogen was obtained by subtracting the C from the 
CH$_2$ total charge changing cross sections using the following relation

\begin{equation}
\sigma_H = \frac{1} {2}(3\sigma_{CH_2}- \sigma_C)
\end{equation}

These measurements were compared with the calculations based on the 
equation (see Table 1)

\begin{equation}
\sigma_{tot(theo)} = \pi r_0^2(A^{\frac{1}{3}}_p + A^{\frac{1}{3}}_t - b)^2
\end{equation}

The experimental total charge changing cross sections increase with 
increasing target mass number. From Table 1 it is seen that for the CH$_2$ 
target the total cross sections increase with the 
increase of the projectile energy.

\begin{figure}[!ht]
\begin{center}
\mbox{\epsfig{figure=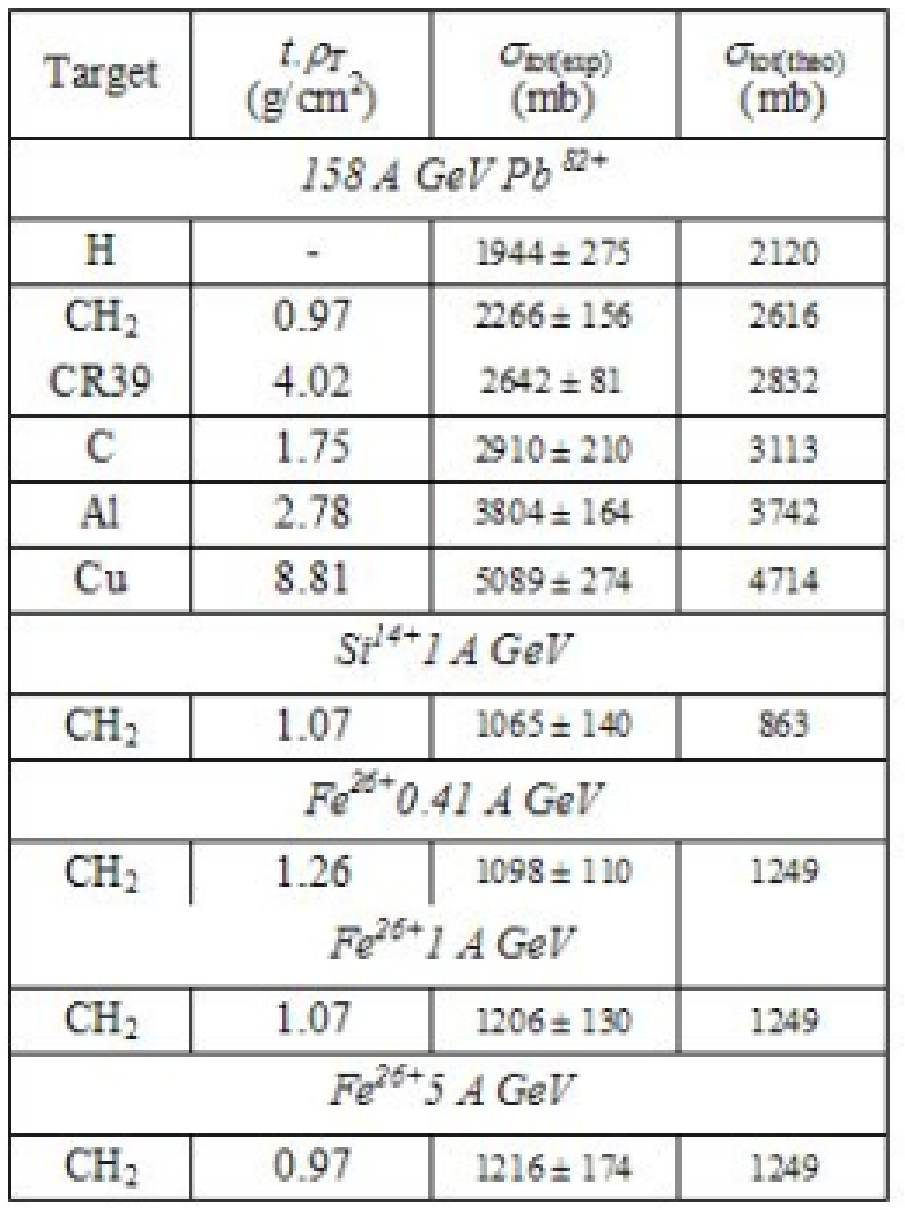,height=7.4cm}}
\label{fig:table1}
\end{center}
\end{figure}

{\normalsize {Table 1: The measured and computed total charge-changing 
cross sections for Pb, Fe and Si projectiles on different targets. The data 
given in the last 4 rows are preliminary. The cross sections on CR39 and
CH$_2$ are averaged on the number of atoms. The quoted uncertainties 
are statistical only.}}

\section{Alpha Radioactivity Measurement}The $\alpha$-radioactivity from 
Po$^{210}$ 
of the OPERA lead plates with 0.07$\%$ Ca or 2.5$\%$ Sb were measured with 
CR39 NTDs and a Surface Barrier Silicon Detector (SBSiD) [16]; comparison of 
the obtained results are shown in Figure 4. We refreshed the CR39 sheets with 
our improved etching conditions before $\alpha$-activity measurements i.e. 
we removed all the pre exposure $\alpha$-activity due to radon or defects in 
CR39 NTDs. The exposure set up is sketched in the inset of Fig. 4.

\begin{figure}
\begin{center}
\mbox{\epsfig{figure=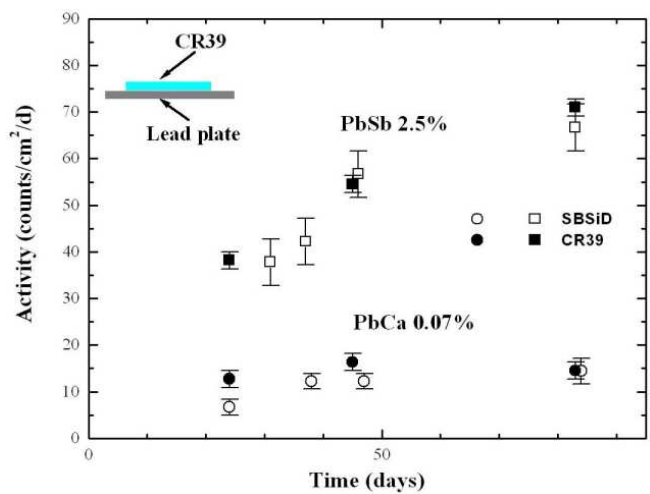,height=6.5cm}}
\caption{Time dependence of $\alpha$-radioactivity in PbCa and PbSb plates measured with CR39 NTDs and the SBSiD.}
\label{fig:4}
\end{center}
\end{figure}

\section{Conclusions}With use of alcohol we have improved the quality of 
the post etched surface, removed background and enhanced the sharpness of 
the tracks in CR39 and Makrofol NTDs, but the thresholds are higher. 
New calibrations curves were obtained with Pb$^{82+}$, In$^{49+}$, 
Fe$^{26+}$ and Si$^{14+}$ relativistic heavy ions and improved measuring 
techniques [17-18].\par
The use of CR39 and Makrofol detectors etched with alcohol can be useful for 
experiments having a substantial background originated by local sources and 
in the case of cosmic ray long duration balloon flights, or space 
experiments.\par
The measured total charge changing cross sections are in agreement with the 
calculation based on eq. 6 within statistical standard deviations. On the 
basis of our results, we conclude that NTDs, especially CR39, can be used 
for high energy experimental cross section studies. \par
The improved etching and analysis methods can be applied for the detections 
of alpha particles from radon and for environmental dosimetry. 
We successfully applied this new procedure for the measurements of alpha 
radioactivity in lead plates.\\

{\bf Acknowledgements.} We thank the CERN SPS, BNL and HIMAC staff for the 
beam exposures. We gratefully acknowledge the contribution of our technical 
staff in Bologna. We thank INFN and ICTP for providing fellowships and 
grants to non-Italian citizens.


\begin{thebibliography}{18}

\bibitem{1} M. Ambrosio et al., Eur. Phys. J. C25 (2002) 511; Nucl. Instrum. 
Meth. A486 (2002) 663.

\bibitem{2} M. Ambrosio et al., Eur. Phys. J. C13 (2000) 453, hep-ex 0009002.

\bibitem{3} A. Kumar et al., Radiat. Meas. 36 (2003) 301.

\bibitem{4} S. Cecchini et al., Astropart. Phys. 1 (1993) 369.

\bibitem{5} S. Cecchini et al., Nucl. Phys. A707 (2002) 513.

\bibitem{6} H. Dekhissi et al., Nucl. Phys. A662 (2000) 207.

\bibitem{7} T. Chiarusi et al., Radiat. Meas. 36 (2003) 335.

\bibitem{8} M. Beozzo et al., Nucl. Tracks Radiat. Meas. 19 (1991) 297.

\bibitem{9} S. Cecchini et al., Nuovo Cimento 24C (2001) 639. 

\bibitem{10} S. Cecchini et al., Nuovo Cimento 109A (1996) 1119. \par
G. Giacomelli et al., Radiat. Meas. 28 (1997) 217.

\bibitem{11} S. Manzoor et al., Nucl. Instrum. Meth. A453 (2000) 525.

\bibitem{12}  H. A. Khan et al., Radiat. Meas. 31 (1999) 25.\par 
G. Giacomelli et al., hep-ex/0005041.

\bibitem{13} A. Noll et al., Nucl. Tracks Radiat. Meas. 15 (1988) 265.

\bibitem{14} S. A. Durrani et al., Solid State Nuclear Track Detection, 
Pergamon press (1987).

\bibitem{15} G. Giacomelli et al, Nucl. Instrum. Meth. A411 (1998) 41.

\bibitem{16} OPERA Int. Note (2006), Private Comm.

\bibitem{17} S. Balestra et al., accepted for publication Nucl. Instrum. Meth. 
B, physics/0610227.

\bibitem{18} V. Togo et al., 10th Inter. Symp. Radiat. Phys., Coimbra, Portugal, 17-22 Sept. 2006: to be published in Nucl. Instrum. Meth. A, physics/0611105.



\end{thebibliography}
\end{document}